\documentclass[fleqn,twoside,a4paper]{article} \usepackage{espcrc2}
\usepackage{graphicx}
\usepackage[figuresright]{rotating}

 \newcommand{\MeV}{\rm MeV}
\newcommand{\GeV}{\rm GeV} 

\title{A precise determination of the $B_c$ mass from dynamical
  lattice QCD} \author{I. F. Allison \address[GLA]{Theoretical Physics
    Group, Department of Physics and Astronomy, University of Glasgow,
    \\ Glasgow G12 8QQ, UK}\thanks{Talk presented by I.  Allison at
    Lattiec 2004, Fermi National Accelerator Laboratory, June 21-26,
    2004}, C.T.H.  Davies\addressmark[GLA], A. Gray
  \address[OHIO]{Physics Department, The Ohio State University,
    Columbus, Ohio 43210, USA}, A. S.  Kronfeld\address[FNAL]{Fermi
    National Accelerator Laboratory, Batavia, Illinois 60510, USA},
  P.B. Mackenzie\addressmark[FNAL],
  J. N. Simone\addressmark[FNAL] \\
  (HPQCD, FNAL lattice and UKQCD collaborations)}
\begin{document}

\begin{abstract}
  We perform a precise calculation of the mass of the $B_c$ meson
  using unquenched configurations from the MILC collaboration,
  including $2+1$ flavours of improved staggered quarks. Lattice
  $\textsc{NRQCD}$ and the Fermilab formalism are used to describe the
  $b$ and $c$ quarks respectively. We find the mass of the $B_c$ meson
  to be $6.304(16) \GeV$.
\end{abstract}

\maketitle

\section{Introduction}
One of the results anticipated from run II~\cite{Anikeev:2001rk} at
the Tevatron is a measurement of the largely unexplored properties of
the $B_c (b\bar{c})$ meson. The current experimental determination
comes from the $CDF$ collaboration who quoted a value of
$6.4(4)~\GeV$~\cite{Abe:1998wi} and from preliminary $D\emptyset$
data, where $231$ events have been identified, giving the $B_c$ mass
as $M_{B_c}=5.95(37)~\GeV$~\cite{Jesik:2004pc}. A suitably precise
lattice calculation puts us in a position to make a prediction of this
mass, rather than confirm an experimental result.

In the past, the two main theoretical methods for finding $M_{B_c}$
have been potential models~\cite{Eichten:1994gt,Kwong:1990am} and
lattice calculations in the quenched
approximation~\cite{Shanahan:1999mv,Davies:1996gi}. Both of these
techniques have found agreement with the experimental values, but both
have drawbacks; potential models inevitably depend on the form of the
potential used, while the effect of quenching in a lattice calculation
is estimated to be about $100~\MeV$.

Now, however, it is possible to repeat the lattice calculation without
the quenched approximation. The key is to use the unquenched ensembles
for lattice gauge fields from the \textsc{MILC}
collaboration~\cite{Bernard:2001av}. These have 2+1 flavours of
dynamical quarks. They use an improved gluon action, and the Asqtad
action for staggered light quarks, leaving discretisation errors of
$O(\alpha_s a^2)$. With the \textsc{MILC} ensembles, lattice
calculations agree with experimental measurements for a wide range of
hadronic quantities~\cite{Davies:2003ik}.

\section{Method}
%
%
%
%
%
In bottomium $(b\bar{b})$ systems the typical velocity of the quarks
is $v_b\approx0.1$, while in charmonium $(c\bar{c})$ $v_c\approx0.3$.
However, the unequal sharing of quark masses in the $B_c$ changes
these values, giving $v_c\approx0.5$ and $v_b\approx0.04$. With this
in mind, we choose to use different formalisms to describe the $b$ and
$c$ quarks.

For the $c$ quark we use the clover action, with the non-relativistic
(Fermilab) interpretation~\cite{El-Khadra:1996mp}. The clover coupling
is adjusted to its tadpole improved tree-level value. For the $b$
quark we use $O(v^4)$ lattice \textsc{NRQCD}~\cite{Davies:1994mp} with
coefficients fixed at tree level. Our reasoning for these choices
comes from the charmonium spectrum, reproduced using Fermilab
quarks~\cite{diPierro:2003bu}, and from the $\Upsilon$ spectrum,
reproduced using the \textsc{NRQCD} action~\cite{Gray:2002vk}. Both of
these calculations were performed on the \textsc{MILC} $2+1$ flavour
ensembles which our calculations use.

In both the \textsc{NRQCD} and Fermilab formalisms, the hadron energy
at zero momentum is offset from the mass by an energy shift. In
appropriate mass differences, this shift cancels. Therefore,
\begin{equation}
  \label{eq:2}
  M_{B_c} -\frac{1}{2}\left[ M_\psi + M_\Upsilon \right] = E_{B_c} -
  \frac{1}{2}\left[ E_\psi + E_\Upsilon \right],
\end{equation}
and
\begin{equation}
  \label{eq:3}
  M_{B_c} -\left[ M_{B_s} + M_{D_s} \right] = E_{B_c} - 
  \left[ E_{B_s} + E_{D_s} \right],
\end{equation}
yield the differences in binding energies. (Here $M_X$ and $E_X$
denote masses and energies calculated on the lattice respectively.)
After computing these differences on the lattice, we obtain our result
for $M_{B_c}$ by adding back $[M_\Psi+M_\Upsilon]/2$ or
$M_{D_s}+M_{B_s}$. These methods for extracting $M_{B_c}$ will be
referred to as the \emph{quarkonium baseline}~(\ref{eq:2}) and the
\emph{heavy-light baseline method}~(\ref{eq:3}). When results are
quoted for the quarkonium baseline method, the ``charmonium particle''
used is the spin average of the $\eta_c$ and $J/\psi$. The absence of
an experimental signal for $\eta_b$ prevents us from doing this for
the $b$ quark as well.

\section{Simulation Details}
In our calculation, the configurations included a single flavour at
around the strange quark mass ($m_s$) and two degenerate light
flavours over a range of masses down to $m_s/5$. The valence $c$ and
$b$ quark masses were set using the kinetic masses of the $D_s$ and
$\Upsilon$ states. The lattice spacing is set on each ensemble through
the 1S--2S radial splitting of the $\Upsilon$ system. But an important
advantage of calculations on the \textsc{MILC} ensembles is that any
of several quantities shown in~\cite{Davies:2003ik} could have been
used.

In addition to the majority of our calculations performed at
$a\approx0.12\,\mbox{fm}$ we ran a single calculation on a finer
ensemble, at $a\approx0.09\,\mbox{fm}$.

\section{Results}
\begin{figure}[htb]
  \centering \includegraphics[width=\linewidth]{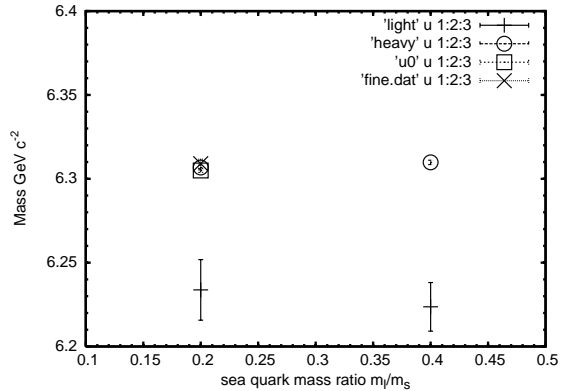}
  \caption{Chiral limit of quarkonium and heavy-light baseline
    methods. Only statistical errors are shown. The $m_l/m_s$ ratio
    corresponds to the bare masses used for the dynamical quarks.}
  \label{fig:chirboth}
\end{figure}

The result of calculating using different masses for the two
(degenerate) light sea quarks is shown in Fig.~\ref{fig:chirboth}.
This plot also allows a comparison of the quarkonium and heavy-light
baseline methods of Eqs.~(\ref{eq:2}) and~(\ref{eq:3}).

We plot our result from calculating at the ``fine'' lattice spacing
($a\approx0.09fm$). We find on these configurations that
$M_{B_c}=6.309(3)~\GeV$, (quoting only the statistical error).  We
also plot a single point where the tadpole improvement factor $(u_0)$
for the $b$ quark has been calculated using the mean link in Landau
gauge as opposed to the fourth root of the plaquette (used for all
other points). Changing only this quantity allows us to assess the
influence of the relativistic corrections in the \textsc{NRQCD}
action. We see that the effect is negligible, as expected.


The effect of varying the valence quark masses on the kinetic masses
of the $J/\psi$ and $\Upsilon$ was used to set the error on the $B_c$
from these parameters. We find that this leads to a error of $10~\MeV$
due to the uncertainty in the $b$ quark mass and $5~\MeV$ due to the
uncertainty for the $c$.

We also estimate a systematic error arising from higher order terms in
the Fermilab quark action.  For the heavy quarks, the \textsc{NRQCD}
action includes the $O(v^4)$ terms at tree level, the Fermilab quark
action also includes them but with some mismatch. Because of the form
of Eqs.~(\ref{eq:2}) and~(\ref{eq:3}) we expect a more efficient
cancellation of these differences using the quarkonium baseline
method, where all quantities are in the heavy quark sector.
Estimating the corrections, we get a $\pm 10~\MeV$ uncertainty for
both methods, and an additional downward shift of $30-50~\MeV$ for the
heavy-light baseline method. Evidence for this shift can be seen in
Fig.~\ref{fig:chirboth}.

We give our final result using the quarkonium baseline method because
of this shift. Our result is $6.304(3)(10)(11)(5)(3)~\GeV$, where the
first error is statistical, the second is due to higher order
corrections to the Fermilab quark action, the third is due to
uncertainties in the valence quark masses, and the final number is an
estimate of our chiral extrapolation error. Further error analysis is
still underway, but combining the current errors in quadrature gives:
\begin{equation}
  \label{eq:1}
  M_{B_c}=6.304(16)~\GeV.
\end{equation}

\begin{figure}[t]
  \centering
  \includegraphics[width=1.0\linewidth]{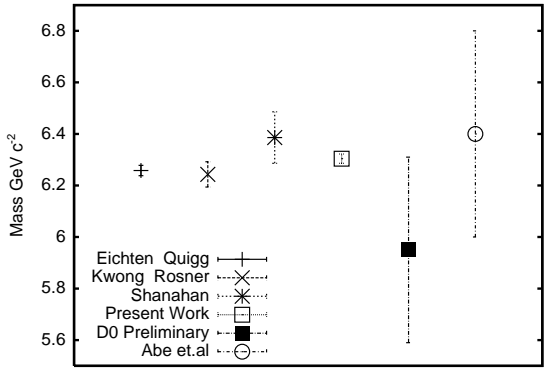}
  \caption{Mass of the $B_c$ from potential models, Quenched lattice
    calculation and the present work (quarkonium baseline method).}
  \label{fig:context}
\end{figure}

This result is shown in context in Fig.~\ref{fig:context}. We also
show points due to Eichten and Quigg~\cite{Eichten:1994gt} and Kwong
and Rosner~\cite{Kwong:1990am} who estimate results from a range of
potentials.

The result due to Shanahan~\cite{Shanahan:1999mv} et al. is the
previous lattice determination. Their calculation used the same method
as this work, but was carried out in the quenched approximation. We
achieve a significant reduction in the size of the systematic error
from the removal of the quenched approximation.

\section{Conclusions}
This calculation shows once again the importance of including
dynamical quarks in lattice simulations. Having removed the ambiguity
in setting the lattice spacing and fixing the quark masses, we are
able to reach an unprecedented accuracy for a lattice calculation of
the mass of the $B_c$ particle. This sets a well defined target for
the current experimental studies.

\section{Acknowledgements}
We gratefully acknowledge the support of PPARC and the DoE.
Particular thanks go to the MILC collaboration for the use of their
configurations. These calculations were performed on the Fermilab
cluster. Fermilab is operated by the Universities Research Association
Inc., under contract with the U.S.\ Department of Energy.


\begin{thebibliography}{10}

\bibitem{Anikeev:2001rk}
K. Anikeev et~al.,
\newblock (2001), hep-ph/0201071.

\bibitem{Abe:1998wi}
CDF, F. Abe et~al.,
\newblock Phys. Rev. Lett. 81 (1998) 2432, hep-ex/9805034.

\bibitem{Jesik:2004pc}
R. Jesik,
\newblock B physics at $\mbox{D}\emptyset$,
\newblock Unpublished talk given at Imperial College London, 2004.

\bibitem{Eichten:1994gt}
E.J. Eichten and C. Quigg,
\newblock Phys. Rev. D49 (1994) 5845, hep-ph/9402210.

\bibitem{Kwong:1990am}
W.k. Kwong and J.L. Rosner,
\newblock Phys. Rev. D44 (1991) 212.

\bibitem{Shanahan:1999mv}
UKQCD, H.P. Shanahan et~al.,
\newblock Phys. Lett. B453 (1999) 289, hep-lat/9902025.

\bibitem{Davies:1996gi}
C.T.H. Davies et~al.,
\newblock Phys. Lett. B382 (1996) 131, hep-lat/9602020.

\bibitem{Bernard:2001av}
C.W. Bernard et~al.,
\newblock Phys. Rev. D64 (2001) 054506, hep-lat/0104002.

\bibitem{Davies:2003ik}
HPQCD, C.T.H. Davies et~al.,
\newblock Phys. Rev. Lett. 92 (2004) 022001, hep-lat/0304004.

\bibitem{El-Khadra:1996mp}
A.X. El-Khadra, A.S. Kronfeld and P.B. Mackenzie,
\newblock Phys. Rev. D55 (1997) 3933, hep-lat/9604004.

\bibitem{Davies:1994mp}
C.T.H. Davies et~al.,
\newblock Phys. Rev. D50 (1994) 6963, hep-lat/9406017.

\bibitem{diPierro:2003bu}
M. di~Pierro et~al.,
\newblock Nucl. Phys. Proc. Suppl. 129 (2004) 340, hep-lat/0310042.

\bibitem{Gray:2002vk}
HPQCD, A. Gray et~al.,
\newblock Nucl. Phys. Proc. Suppl. 119 (2003) 592, hep-lat/0209022.

\end{thebibliography}
\end{document}